\documentclass[prb,showpacs,amssymb,amsfonts,twocolumn,superscriptaddress]{revtex4}

\usepackage{amsmath,amsthm,amsfonts,graphicx,epsfig}
\usepackage[usenames]{color}
\usepackage{bm}

\newcommand{\dbar}{\kern-.1em{\raise.8ex\hbox{ -}}\kern-.6em{d}}

\def\half{\mbox{$1\over2$}}

\def \be{\begin{equation}}
\def \ee{\end{equation}}
\def \beq{\begin{equation}}
\def \eeq{\end{equation}}
\def \barray{\begin{eqnarray}}
\def \earray{\end{eqnarray}}
\begin{document}
\title{Spontaneously localized photonic modes due to disorder in the dielectric constant}
\author{Y. Kodriano}
\author{ D. Gershoni}
\email{dg@physics.technion.ac.il}
\author{ E. Linder}
\author{ B. Shapiro}
\affiliation{Department of Physics, Technion---Israel Institute of
Technology, 32000 Haifa, Israel.}
\author{M. E. Raikh}
\affiliation{Department of Physics, University of Utah, Salt Lake
City, Utah 84112, USA.}

\author{J.~P.~Reithmaier}
\affiliation {Technische Physik, Institute of Nanostructure
Technologies and Analysis, Universit\"at Kassel, Heinrich-Plett-Str.
40, 34132 Kassel, Germany.}
\author{S. Reitzenstein},
\author{A. L\"offler},
\author{A. Forchel}
\affiliation {Technische Physik, Universit\"at W\"urzburg, Am
Hubland, D-97074 W\"urzburg, Germany.}
\begin{abstract}
We present the first experimental evidence for the existence of
strongly localized photonic modes due to random two dimensional
fluctuations in the dielectric constant. In one direction, the modes
are trapped by ordered Bragg reflecting mirrors of a planar, one
wavelength long, microcavity. In the cavity plane, they are
localized by disorder, which is due to randomness in the position,
composition and sizes of quantum dots located in the anti-node of
the cavity. We extend the theory of disorder induced strong
localization of electron states to optical modes and obtain
quantitative agreement with the main experimental observations.
\end{abstract}

 \maketitle

\section{introduction}
Localization of a photonic mode by disorder in a statistically
uniform and isotropic medium has been studied for more than two
decades~\cite{wiersma,sheffold,chabanov,laurent}, both in the
optical and microwave spectral ranges. In this paper we propose and
demonstrate a new mechanism for
disorder induced strong three dimensional (3D) localization of
light. We achieved this in planar microcavities (PMCs), containing a
layer of strain-induced self-assembled quantum dots
(QDs)~\cite{vahala,guy,sanvitto}. The dispersion of light in the PMC
plane is modified by the distributed Bragg reflectors (DBRs) that
trap the modes in perpendicular direction to the cavity plane. The
QDs constitute an active material which, under proper excitation,
emits photons, due to recombination of QD confined
excitons~\cite{gerard,akopian:2006}. Randomness in the QDs in-plane
position, composition and sizes produces fluctuations in the
system's dielectric constant. This randomness provides the necessary
``attractive potential'', which, together with the modified
dispersion, brings about strong photon localization in the PMC
plane. Our novel mechanism differs from previous studies, since it
provides genuine disorder-induced localization rather than leaky
(resonant) modes~\cite{wiersma,sheffold,chabanov,laurent}. It also
differs from the ``transverse localization''~\cite{de_raedt}, where
the wave freely propagates in one direction while being trapped, by
disorder, in the transverse direction~\cite{schwartz}. We directly
demonstrate the 3D localization of the photonic modes by measuring
their photoluminescence (PL) intensity distribution.

It is worthwhile to emphasize from the start that the comparatively small size
of the excitation spot ($30 \ \mu m$, or smaller) plays a crucial role in the
interpretation of the experimental data. The point is that \textit{in the
absence of disorder} the shape of the PL spectrum emerging from the cavity is
quite sensitive to the size of the excitation spot. If the size is sufficiently
small, the PL spectrum becomes very narrow, in marked difference with the broad
shoulder predicted by the standard 'basic cavity physics'. If there had been no
disorder, the experimental conditions of our paper would have resulted in an
extremely narrow PL peak. The significant broadening of the spectrum, as
observed in our experiment, is due entirely to disorder. The broadening occurs
both towards frequencies below and above the cavity frequency (localized and
resonant photonic modes, respectively). All this will be discussed in detail in
the theoretical part of the paper.

\section{Experimental}

\begin{figure}[htbp]
\epsfxsize=.52\textwidth\centerline{\epsffile{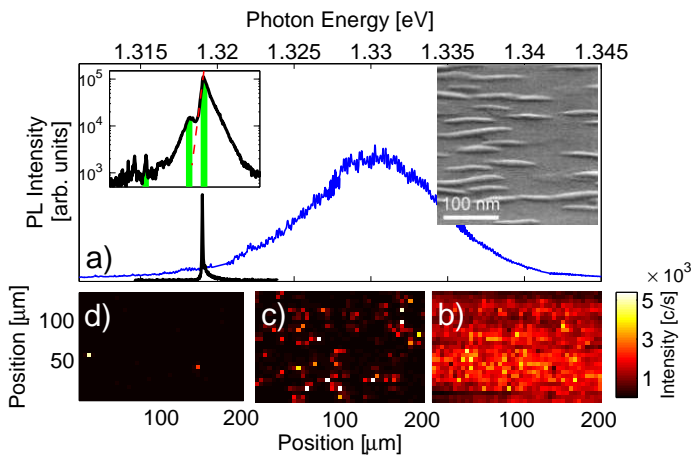}}
\vspace*{-0.4cm} \caption{ a) PL spectrum of the PMC with (solid
black line) and without (solid blue line) the upper DBR. The left
inset represents spatially integrated PL spectrum from a 135$\mu m
\times$200$\mu m$ area on the sample surface. b)-d) selective
wavelength images of the area for 3 different spectral domains
indicated by the green zones in a). The right inset is a scanning
electron micrograph of the QDs layer. } \label{fig:one}
\end{figure}

Our sample was grown by molecular beam epitaxy on a [1,0,0] oriented
GaAs substrate. The sample consists of a PMC formed by a GaAs
one-wavelength resonator, sandwiched between DBRs made from 23(26)
top (bottom) alternating GaAs and AlAs quarter-wavelength-layer
pairs. As an active material in the resonator's anti-node, a single
strain-induced self-assembled In$_{0.3}$Ga$_{0.7}$As QDs layer with
average density of 5$\times 10^9$ QDs per $ cm^{2} $ is
used~\cite{ref1}. The use of such composition results in large and
asymmetric QDs with typical lengths and widths of about 100 and 30
nm respectively, as can be seen in the inset to Fig.~\ref{fig:one}a.

We used two methods to study the PMC. a) a diffraction limited
confocal optical scanning microscopy ( a $\times$60 objective with
numerical aperture of 0.85, obtaining spatial resolution of $\sim
0.7 \mu m$). b) a near field scanning optical microscopy (NSOM-
Nanonics Cryoview $2000^{TM}$ with spatial resolution of $\sim 0.25
\mu m$). By applying either method we measure the lateral
distribution of the electromagnetic field above the sample surface.
Two modes of excitation were used. In the first, the excitation was
focused to a diffraction limit by the collecting objective
~\cite{ref2}. In the second (the only one used for NSOM), the
excitation was focused at an oblique angle to a spot diameter of
about 30 $\mu m$. Due to the high energy used for the excitation
(632.8 nm light of a HeNe laser) the images produced by the two
excitation modes were almost identical. In Fig.~\ref{fig:one}a we
present PL spectrum (upper, black solid line) from a single point on
the surface of the PMC sample together with a PL spectrum which was
taken from the same sample (lower, solid blue line) after the upper
DBR mirror was completely etched. The spectrally broad blue PL line
is centered at 1.33 eV and has full width at half maximum (FWHM) of
approximately 10 meV. This spectral line results from s-shell
electron-hole recombination within the inhomogeneous population of
QDs of various sizes and compositions. The radiative linewidth of a
QD is $\sim 2 \mu eV$~\cite{akopian:2006}.

The rather symmetrical shape of the spectral line indicates that
charge carriers are not distributed thermally between the QDs.
Therefore, it is quite safe to assume that the spectrum accurately
represents the actual energy distribution of the emission from
individual QDs. The spectrally sharp PL line in the figure is
centered at 1.319 eV and has FWHM of 80 $\mu$eV. As we explain
below, in a disorder free PMC this line width should have been about
an order of magnitude narrower due to the finite excitation area and
diffraction limited collection spot.

The asymmetric spectral shape of the line and its actual linewidth
which represents a Q-factor of $\sim$15000, in similarity to earlier
measurements~\cite{sanvitto,vahala,guy},  are due to the disorder in
the dielectric constant as we quantitatively show below.

 In the left inset to Fig.~\ref{fig:one}a we present on a
semi logarithmic scale a spatially integrated (`far field') PL
spectrum from a 135$\mu m$ wide $\times 200 \mu m$ long area on the
sample's surface. The spectrum was obtained by summing up large
number of individual PL spectra, each obtained from a diffraction
limited areal spot. The dominating spectral feature is a large,
asymmetrically broadened line around the energy of the PMC mode. The
line decays exponentially towards lower energies with a
characteristic energy of $150\pm 30 \mu eV$. Towards higher energies
the line decays with a characteristic energy of $380\pm 30 \mu eV$.
The ratio between these characteristic energies agrees well with the
prediction of our theoretical model for the disorder in the
dielectric constant (see below).

In Figs.~\ref{fig:one}b-\ref{fig:one}d selective wavelength images
of the scanned area are presented. The spectral domains which are
used for each image are indicated by the green zones imposed on the
'far field' spectrum in Fig.~\ref{fig:one}a. The image in
Fig.~\ref{fig:one}b is obtained within a 0.5 meV energy window
containing the PMC mode. It shows almost evenly distributed emission
from the surface. In contrast, the image in Fig.~\ref{fig:one}c,
which is obtained within a 0.5 meV window, 1 meV below the PMC mode,
shows emission emerging from randomly distributed spots on the
surface. Likewise, the image in Fig.~\ref{fig:one}d, which is
obtained from energy window of same width, located 3 meV lower,
shows only 2 bright centers, from which all the emission results.

\begin{figure}[htbp]
\epsfxsize=.46\textwidth \centerline{\epsffile{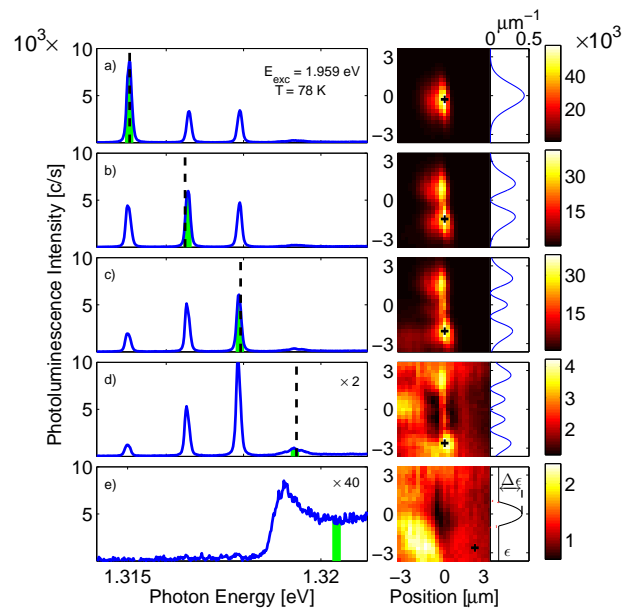}}
\vspace*{-0.4cm}
\caption{ a)-e) middle panels - selective wavelength NSOM images of an area in
the vicinity of the lowest energy mode observed in Fig. \ref{fig:one}.
The spectra (left panels) are obtained from the brightest
area pixels indicated by the + signs on the images.
The green areas on the spectra, indicate the spectral domains used for
generating the images. The right panels represent analytically
calculated 1D intensity distributions of the confined optical modes
obtained by solving Eq.~(\ref{eq:two1}) with the parabolic
dielectric constant fluctuation model presented in the right panel
of e) where $\Delta \epsilon/\epsilon = 0.7\%$ and $\epsilon$ the
dielectric constant of GaAs. The calculated mode energies are
presented by dashed vertical lines on the spectra to the left.}
\label{fig:two}
\end{figure}

In Fig.~\ref{fig:two} we turn our attention to one of the low energy
electromagnetic modes seen in Fig.~\ref{fig:one}d. Such localizing
centers were observed in various mode energies and various locations
on the sample surface. In the right panels of
Figs.~\ref{fig:two}a-\ref{fig:two}e, we display selective wavelength
images obtained by scanning the Aluminum coated fiber tip of our
NSOM microscope in tapping mode over the sample surface. The
corresponding spectra in the left panels, are obtained from the
brightest area pixels indicated by the + signs on the images. The
green areas on the spectra indicate the spectral domains used for
generating these images. These spectra and corresponding images
reveal the following observations: Four different localized photonic
modes are observed at energies below that of the PMC mode. Each mode
is composed of two cross linearly polarized equally intense
components (not shown). The components are $\sim 80 \mu eV$ apart
and the lower energy one is polarized along the [0,-1,1]
direction~\cite{bacher}. The lowest energy mode is about 5 meV below
the energy of the PMC mode. The in-plane spatial intensity
distribution of this mode, as depicted by the image in
Fig.~\ref{fig:two}a, is node-less, and has an elongated shape
oriented along the [0,-1,1] crystallographic direction, along which
the QDs are also elongated (see inset to Fig.~\ref{fig:one}). The
mode dimensions are roughly 3$\times$2$\mu m^2$. The peak intensity
is about 40 times stronger than that of the PMC mode. The next three
images in Fig.~\ref{fig:two}b, c and d, present elongated spatial
intensity distributions composed of one, two and three nodes,
respectively. Each mode's energy is $\sim$1.5-2 meV higher than the
energy of the preceding mode. The selective wavelength image in
Fig.~\ref{fig:two}e is obtained at an energy window, which resides
on the higher energy side of the PMC mode. It displays evenly
distributed emission from everywhere except from the actual locus of
the localized modes, where extended modes are forbidden. The
magnitude and shape of the spot from which emission at this
wavelength is missing is an estimate for the spatial extent of the
region which localizes the optical modes. The region is comparable
in shape and size to the most localized, lowest energy mode. The
localized modes are insensitive to temperature and excitation
density variations, hence they do not result from thermal or
non-linear effects. The emission is linear with the excitation
density between 1-$10^6$ Watt/$cm^2$. Increasing the temperature
from 10 to 80K results in a small, rigid, red spectral shift, as
expected from the temperature dependence of the material bandgap.

\section{Theory}
Since the sizes of the spots in Fig. 2b-2d significantly exceed the
size of an individual QD, the only way to account for the
experimental observations is to assume that each spot corresponds to
a {\em localized} electromagnetic mode. Therefore the low frequency
tail of the PL spectrum in Fig. 1a should be interpreted as a tail
of localized modes with energies below that of the PMC mode,
$\omega_c$, similar to the tail of localized electron states formed
below the band-edge due to disorder~\cite{ref3,ref4}. To demonstrate
how the presence of a cavity leads to the in-plane trapping, we
consider the scalar wave equation corresponding to the frequency
$\Omega$: \beq\label{eq:one1} {d^2\Psi \over dz^2} +
\nabla^2_{\bm{\rho}} \Psi + {\Omega^2 \over c^2} \Bigl[
\varepsilon_0 + \delta \varepsilon(\bm{\rho},z) \Bigr]\Psi = 0, \eeq
where $\varepsilon_0$ is a uniform dielectric constant inside the
cavity, $\bm{\rho} \equiv (x,y)$, and $\delta
\varepsilon(\bm{\rho},z)$ is the perturbation due to the randomly
positioned QDs that reside in a narrow layer, of width $\delta z
\sim 5$ nm, in the middle of the cavity. Let us emphasize that an
individual QD is much too small to be able to trap a photon.
Actually, in the continuum description employed in
Eq.~(\ref{eq:one1}) individual QDs are not resolved. What matters
are the fluctuations in the areal density of the QDs, which manifest
themselves in fluctuations of the dielectric constant $ \delta
\varepsilon $. Since the electromagnetic modes are confined in the $z$-direction
to a much shorter length than in the plane, the function
$\Psi(\bm{\rho}, z)$ can be factorized as $\Phi(\bm{\rho})
\textrm{cos} (\pi z / d)$, where $d$ is the cavity width, and
$\Phi(\bm{\rho})$ satisfies the equation \beq\label{eq:two1} -  c^2
\nabla^2_{\bm{\rho}} \Phi - {\Omega^2 } \delta
\varepsilon(\bm{\rho}) \Phi = - {\varepsilon_0} (\omega_c^2 -
\Omega^2) \Phi. \eeq In this equation,
$\delta\varepsilon(\bm{\rho})$ is the effective in-plane fluctuation
of the dielectric constant obtained from
$\delta\varepsilon(\bm{\rho},z)$ by averaging over the
$z$-distribution of the cavity mode intensity, namely,
$\delta\varepsilon(\bm{\rho})= \frac{2}{d}\int_{-d/2}^{d/2}\delta
\varepsilon(\bm{\rho},z)\cos^2(\pi z/d)$. The analogy between Eq.
(\ref {eq:two1}) and the Schr\"odinger equation makes it clear that,
for positive $\delta \varepsilon(\bm{\rho})$ , Eq.~(\ref{eq:two1})
admits localized solutions, with frequency $\Omega$ smaller than the
cavity frequency $\omega_c = 2 \pi c / d \sqrt{\varepsilon_0}$.
Thus, even a small enhancement in $\delta \varepsilon(\bm{\rho})$
acts as an attractive potential localizing the mode laterally. To
substantiate that this mechanism of trapping is relevant to our
experimental observations, we note that the spatial intensity
distributions and the energies of the modes in Fig.~\ref{fig:two}
are quite accurately fitted by solving Eq.~(\ref{eq:two1}) using a
one dimensional parabolic model $\Bigl(
 \delta \varepsilon (x)=\Delta \varepsilon [1-( \alpha x)^2] $,
$| \alpha x
 | \le 1 \Bigr)$
describing a local fluctuation in the dielectric constant (see
Fig.~\ref{fig:two}e, right panel, where the best fitted values are
$\alpha = 0.3 \mu m^{-1}$ and $\Delta \epsilon/\epsilon_0=0.7\%$).
This agreement, and the fact that even in the case of the most
localized modes, differently polarized components have almost the
same energy, justifies the use of the scalar approximation.

To find the shape of the tail of the localized modes distribution,
we have to specify the disorder ``potential'',
$\delta\varepsilon(\bm\rho)$. The simplest possible model is the
``white-noise potential'' which amounts to treating the QDs as
point-like resonant units, randomly distributed in the plane (this
simple model does not describe the anisotropy observed in the
inset of Fig.~\ref{fig:one} as discussed below). Fluctuations in
the local dielectric constant are given as
\beq\label{eq:three1}
\delta \varepsilon (\bm{\rho}) = A
\int {\delta N_\omega (\bm{\rho}) \over \omega^2 - \Omega^2}, \eeq
where $\delta N_\omega(\bm{\rho})$ is the deviation of the number
of QDs (per unit area near $\bm{\rho}$), with resonance frequency
in the interval $d\omega$, from the average value $\bar N P_D(\omega) d\omega$. Here
$\bar N$ is the total areal density of the QDs, regardless of
their frequency and $P_D(\omega)$ is the probability distribution
of the resonance frequencies (the solid, blue line in Fig.
\ref{fig:one}).

To estimate the constant $A$, we consider the $\Omega \rightarrow 0$ limit
and imagine that the entire layer consists of QDs, i.e. of
In$_{0.3}$Ga${_{0.7}}$As. In this case, the enhancement of the
dielectric constant in the layer, as compared to the surrounding
material of GaAs, is $(\varepsilon_{InAs} - \varepsilon_{GaAs})
\cdot 0.3$. This value should be multiplied by $\delta=5/300$, which
is the ratio of the QDs layer width (5 nm) to the cavity width
($\sim$300 nm). There is an extra factor of 2, since the QDs are
located at the anti-node of the PMC mode. Thus, the effective
enhancement, due to the presence of InAs, is $\Delta \varepsilon
\cong
 2.25 \times 10^{-2}$. The constant $A$ can now be estimated
from Eq.~(\ref{eq:three1}) as $A
\cong 2.25 \times 10^{-2} S \omega_0^2$, where $S$ is the typical
area of a QD and $\omega_0 \approx \omega_c$ is the central
frequency of the distribution $P_D(\omega)$. Note that the employed
averaging in the $z$-direction, as well as our use of the scalar
approximation is justified by the smallness of the parameter
$\delta$. Indeed, the effect of disorder-induced coupling between different
cavity modes is proportional to $\delta^2$, whereas
the effect of the QD layer on a
given mode is proportional to $\delta$.

The function $\delta N_\omega (\bm{\rho})$ is a complicated, random
function of position $\bm{\rho}$ and frequency $\omega$. Using the
white-noise model for $\delta N_\omega (\bm\rho)$ and
Eq.~(\ref{eq:three1}), we obtain the probability $P \left\{ \delta
\varepsilon (\bm{\rho}) \right\}$ for a particular realization
$\delta \varepsilon (\bm{\rho})$ of the fluctuating part of the
dielectric constant
\begin{equation}
\label{eq:four1}
P\left\{\delta \varepsilon \left( \rho \right) \right\} \sim
\textrm{exp} \left\{ - {2 \omega_c^2
\over \bar N A^2} \left( \int {P_D(\omega) d \omega \over (\omega
- \Omega)^2 }\right)^{-1}\!\! \int d^2 \bm{\rho}~[\delta \varepsilon
(\bm{\rho})]^2 \right\}.
\end{equation}
Formally, the integral over $\omega$ in Eq. (\ref {eq:four1}) diverges at
$\omega \approx \Omega$.
In this regard, it should be noted that the integral over $\omega$ is
actually an approximation for a sum over discrete values,
$\omega_i$, of the resonant frequency of the dots in the relevent area (of
order 1~$\mu m^2$).
 The meaning of this area
can be traced back to Eq.~(\ref{eq:three1}) which defines a
macroscopic (albeit still fluctuating in space) dielectric
constant. The spatial scale, over which $\delta \varepsilon (\bm{\rho})$
 is changing, should be at least few times larger than the
``microscopic" distances (size of the dots and their separation,
of the order of 0.1 $\mu m)$. Thus, the minimal scale at which
$\delta \varepsilon (\bm{ \rho})$ can be meaningfully defined
should be in the sub micron range. We are interested in $\Omega$
below $\omega_c$, i.e. in frequency which belongs to the tail of
the distribution $P_D (\omega)$. A simple estimate shows that for
such $\Omega$ the probability to encounter a dot with $\omega_i$
close to $\Omega$ is extremely small.
Moreover, for such $\Omega$ the integral over
$\omega$ in Eq.~(\ref{eq:four1}) can be, roughly, replaced by
$(\omega_0 - \Omega)^{-2} $ (note that such a replacement would
not be possible for $\Omega$ in the ``bulk" of the probability
distribution $P_D (\omega)$).
The two equations, (\ref{eq:two1}) and
(\ref{eq:four1}), define the statistical problem of
finding the probability that a localized mode,
in a given frequency interval, will be created.
This problem is completely identical to the
two-dimensional version~\cite{ref4g} of the problem
of the tails of electron states in a random
potential~\cite{ref3,ref4}. The
final result for the probability ${\cal P}(\Omega) d \Omega d S$
to find a trap, in an area $dS$, which can support a
localized mode in the
frequency range $d \Omega$ can
be presented as
 \beq\label{eq:five1}
{\cal P}(\Omega) \cong {\varepsilon_0 \omega_c \over \pi c^2}
\textrm{exp} \left[ - 32 \pi {\varepsilon_0 c^2 \over A^2 \bar N}
{\omega_c - \Omega \over \omega_c} \cdot(\omega_0 - \Omega)^2
\right],
 \eeq
where the pre-exponential factor
corresponds to the density of the PMC
modes at frequencies $\Omega
\gtrsim \omega_c$.
Eq.~(\ref{eq:five1}) can be viewed as a qualitative interpolation
between the maximal value of the modes density at $\Omega =
\omega_c$, and the low-density ``tail''.
A more detailed crossover behavior between
the tail at $\Omega <\omega_c$ and ${\cal P}=\varepsilon_0 \omega_c/\pi c^2={\cal P}_0$
at $\Omega \gtrsim \omega_c$ can be found in Ref.~\onlinecite{Thouless}.
Important is that this crossover is determined by a {\em single} scale of
frequencies
\begin{equation}
\label{omegat}
\omega_t=\omega_c\frac {A^2 {\bar N}}{32 \pi
\varepsilon_0  c^2(\omega_0 - \omega_c)^2},
\end{equation}
which follows from Eq. (\ref{eq:five1}), so that
in the tail region
${\cal P}(\Omega) $ can be rewritten
as  ${\cal P}(\Omega)\approx {\cal P}_0\exp\left[-(\omega_c-\Omega)/\omega_t\right]$.
In Eq. (\ref{omegat})  we
replaced $(\omega_0-\Omega)$ by $(\omega_0 - \omega_c)$,
since, experimentally, the difference $(\omega_0 - \omega_c)$ is much
bigger than the tail width, $\omega_t$.

With the above estimate for the constant $A$,
Eq.~(\ref{eq:five1}) yields  a characteristic energy of $100 \pm 70
\mu eV$ for the initial decay of the PL intensity below
$\hbar\omega_c$. It agrees well with the measured value (dashed red
line in the left inset to Fig.~\ref{fig:one}a). The relatively large
uncertainty results from the uncertainties in the QDs density and
their average area.

Up to now we have focused on explaining the PL spectrum tail for $\omega <
\omega_c$. In this domain, the PL intensity, $I(\Omega)$, outside the cavity
is simply proportional to the density of trapped modes, ${\cal P}(\Omega)$. In
other words, the $\Omega$-dependence of the light intensities inside and outside
the cavity is the same for $\omega < \omega_c$.
This is {\em not the case} for the domain of  frequencies $\omega > \omega_c$
to which we now turn. To calculate $I(\Omega)$ in this domain, we
start by recapitulating the basic microcavity physics, in the absence of
disorder. In this case the emission spectrum is controlled by the Airy
factor~\cite{Airy}
\begin{equation}
\label{Airy}
Ai(\Omega,\theta)=\frac{T}
{\vert 1-
r_1r_2\exp\left[2i\phi(\Omega,\theta)\right]\vert^2},
\end{equation}
where $\theta$ is the angle at which the radiation exits the cavity, $\Omega >
\omega_c$ is the radiation frequency, $r_1$ ($r_2$) is the reflection amplitude
from the upper (lower) Bragg mirror and $T = 1 -r_1^2$ is the transmission
coefficient ($r_1$ and $r_2$ are taken to be real). The phase
$\phi(\Omega,\theta) =\sqrt{\varepsilon_0}\Omega d \cos\theta/c$ is acquired in
course of propagation between the boundaries $z=\pm d/2$. At frequencies,
$\Omega$, close to $\omega_c$ Eq. (\ref{Airy}) has a sharp maximum at
$\theta=\theta_{\Omega}$, where the angle $\theta_{\Omega}$ is defined as
\begin{equation}
\label{theta}
\theta_{\Omega}=\left[\frac{2(\Omega-\omega_c)}{\omega_c}\right]^{1/2},
\end{equation}
so that $Ai(\Omega,\theta)$ can be simplified to
\begin{equation}
\label{simplified}
Ai(\Omega,\theta)=\frac{T}
{T^2+4\pi^2(\theta^2-\theta_{\Omega}^2)^2
}.
\end{equation}
Note that condition $\theta=\theta_{\Omega}$
coincides with the  dispersion law of
the waveguided mode. This follows from Eq. (\ref{eq:two1})
in the absence of disorder. Indeed, for in-plane wave vector,
$q=\sqrt{\varepsilon_0}\Omega\sin\theta/c$, Eq. (\ref{eq:two1})
yields $q\approx \sqrt{2\varepsilon_0\omega_c(\Omega-\omega_c)}/c$,
leading to $\theta=\theta_{\Omega}$ for $\theta_{\Omega} \ll 1$.

It follows from Eq. (\ref{simplified})
that the emission spectrum, $I(\Omega)$, has a form
of a plateau, which corresponds to the
basic microcavity physics.
Indeed, integration over $\theta$ yields
\begin{equation}
\label{plateau}
I(\Omega) \propto
\int\limits_0^{\infty}
 d\theta~ \theta Ai(\Omega,\theta)
 = \frac{1}{8}
\Biggl[1+\frac{2}{\pi}\arctan\Bigl(\frac{2\pi\theta_{\Omega}^2}
{T}\Bigr)\Biggr].
\end{equation}
The second term describes a slight smearing of the
plateau near $\Omega=\omega_c$.

Let us consider now photons which exit at an angle $\theta$, from some small
collection spot. It is important to realize that there is a correlation
between the angle $\theta$ and the average distance $L(\theta)$
(measured from the collection spot) at which the corresponding photons
 have been created. Indeed, a typical photon will bounce $1/T$ times before
escaping the cavity. Therefore photons that bounce at angle $\theta$
and escape at the collection spot must be created at a distance
\beq\label{L_theta} L(\theta) \sim T^{-1} d \tan\theta \approx
\frac{\theta d}{T}. \eeq Thus, the finite size of an
\textit{excitation} area can serve as an efficient cutoff for a
maximal escape angle $\theta_m$. Assuming a circular excitation
area, of radius $R$, and choosing a small collection spot at the
center of the circle, one finds a maximal escape angle
\beq\label{theta_m} \theta_m = \frac{RT}{d}. \eeq Of course, this
relation is valid only if $\theta_m$ (and therefore $R$) is not too
large, namely, it should be smaller than the maximal escape angle
for a fully excited cavity. Since the angular cutoff translates into
a frequency cutoff $\Omega_m = \omega_c (1+\half \theta_m^2)$ (see
Eq.~(\ref{theta})), we arrive at the important conclusion that the
frequency range of the collected photons can become very narrow, if
the excitation area is small enough. In our experiment
$T=0.44\times10^{-4}$, $R\cong30\mu$, so that the collected
radiation constitutes a narrow peak at near $\omega_c$, of relative
width \beq\label{peak_width}
\frac{\Omega_m-\omega_c}{\omega_c}=\frac{\theta_m^2}{2}=\frac{1}{2}\Biggl(\frac{RT}{d}\Biggr)^2
\approx 1\times10^{-5}. \eeq Obviously, this peak can be replaced by
a $\delta$-function \beq\label{I0} I(\Omega)=I_0\delta(\Omega -
\omega_c). \eeq Qualitatively, the meaning of Eq. (\ref{I0}) is
that, due to the finite excitation spot, only the photons emitted
normally to the mirror emerge outside the cavity. Quantitatively we
note that the relative width of the experimentally measured PMC mode
($\approx 1\times10^{-4}$) is about an order of magnitude wider than
the one calculated by Eq. (\ref{peak_width}). As we show below, this
broadening and the actual spectral shape of the PMC mode, results
from the disorder. The replacement of the step-like spectrum of the
{\em clean} cavity (for infinite excitation spot) by the
$\delta$-function Eq. (\ref{I0}) (for finite excitation spot) does
not affect the emission spectrum of {\em disordered} cavity in the
domain $\Omega < \omega_c$. In this domain the dependence,
$I(\Omega)$, is still determined by Eq. (\ref {eq:five1}). However,
in the domain $\Omega
> \omega_c$, the $\delta$-peak for the clean cavity gets broadened
due to scattering of {\em extended modes} by the disorder.

In the presence of disorder, the above reasoning, which is based on
the ray picture, is not applicable since {\em all}  the solutions,
$\Phi_{\mu}({\bm \rho})$, of Eq. (\ref{eq:two1}) corresponding to
the eigenstates, $\Omega_{\mu}$, have components with  ${\bm q}=0$
(in the ray picture they correspond to rays with normal incidence to
the microcavity plane). The amplitude of this component is is given
by $\int\!d^2{\bm \rho}~\Phi_{\mu}({\bm \rho})$. Therefore, the
shape of the spectrum acquires the form $I(\Omega)=I_0{\cal
A}(\Omega)$ where the function ${\cal A}(\Omega)$ is defined as
\begin{equation}
\label{I1}
{\cal A}(\Omega)=\sum_{\mu}\int\! d^2{\bm \rho}~\vert\Phi_{\mu}({\bm \rho})\vert^2
\delta(\Omega-\Omega_{\mu}).
\end{equation}
We note that the function ${\cal A}(\Omega)$ as defined by Eq.
(\ref{I1}) is normalized
\begin{equation}
\label{I2}
\int_{-\infty}^{\infty}d\Omega {\cal A}(\Omega)=1,
\end{equation}
Thus, the disorder broadens the $\delta$-peak  (Eq. (\ref{I0}))
without changing its area. Note also, that the calculation of the
spectral shape, ${\cal A}(\Omega)$, is formally reduced to the
calculation of the spectrum of two-dimensional {\em exciton}
absorption in the presence of a white-noise random potential. This
calculation was done earlier in Ref. \onlinecite{shape}, where the
zero-momentum components of the {\em exciton} states were selected
by the matrix element for light absorption~\cite{shape}.

It is also important to note that the spectral shape, ${\cal
A}(\Omega)$, is a function of the ratio
$(\Omega-\omega_c)/\omega_t$, where $\omega_t$ is the same
characteristic frequency, (Eq. (\ref{omegat})), which determines the
spectral shape of the tail of the localized photon states. Indeed,
deep in the tail, $(\omega_c-\Omega)\gg \omega_t$, all the trapped
solutions, $\Phi_{\mu}({\bm \rho})$, at a given
$\Omega_{\mu}=\Omega$ are identical to each other, and Eq.
(\ref{I1}) is essentially reduced to the density equation  of the
trapped modes $\propto \exp\left[-(\omega_c-\Omega)/\omega_t\right]$
(Eq. (\ref{eq:five1})).

In the opposite limit $(\Omega-\omega_c) \gg \omega_t$ the density
of photon states is constant. Here, therefore, the decay of ${\cal
A}(\Omega)$ is determined by the fall-off of the integral
$\int\!d^2{\bm \rho}~\Phi_{\mu}({\bm \rho})$. This integral can be
estimated to first order by  perturbation expansion in
$\delta\varepsilon({\bm \rho})$. Thus it is proportional to
$(\Omega-\omega_c)^{-1}$. Accurate calculation\cite{shape} yields
${\cal A}(\Omega) \approx \omega_t/(\Omega-\omega_c)^2$. This way,
the width of the emission spectrum was found to be $3.7\omega_t$.

The overall spectral shape of ${\cal A}(\Omega)$ was found in Ref.
\onlinecite{shape} by combining the normalized (Eq. (\ref{I2}))
low-frequency and high-frequency asymptotes. The calculated full
width at half-maximum (FWHM) of the emission spectrum is
$3.7\omega_t$,  in excellent agreement with the experiment, where it
is measured to be $3.5\omega_t$. The calculated ratio between the
high energy and low energy slopes is 2:1 in very good agreement with
the measured ratio of ~2.5.

\section{Discussion}

While the white noise model is successful in explaining the initial
drop of the density of localized modes, it cannot account for the
experimental data in the deeper tail, starting around $\Omega =
1.318$ eV and below. It does not explain the non-monotonic decay of
the probability as seen in Fig.~\ref{fig:one}a. It fails in
predicting the observed multiplicity of confined modes which belong
to the same localization region (Fig.~\ref{fig:two}), and it does
not explain the anisotropy in the electromagnetic mode shapes.

The failure of the white noise model, for $\Omega < \omega_c$, is
not surprising. Indeed, when the mode frequency becomes smaller,
localization of the mode becomes tighter. Thus, various local
features of the random potential - such as correlations in
$\delta\varepsilon({\bm \rho})$, the ``granular'' structure of the
disorder and the inherent anisotropy of the system - become more
important. Obviously, all these features are not captured by the
entirely featureless, universal white noise model.

Proper account of these specific features of randomness might
explain some of the data for the deeper $\Omega$-tail. First, once
correlations are introduced, the most probable deep fluctuation is
no longer a fluctuation which captures one mode only. Second, the
probability of finding a localized mode common to two localizing
centers, rather than to one, may prevail at a given energy below the
energy of the PMC mode. Indeed, localized modes between 1.317 eV and
1.318 eV always exhibit a spatial structure consisting of two bright
spots separated by a dark region. This observation suggests
existence of traps of a double-well shape. The origin of such traps,
as well as the multimode traps in the deep tail (below 1.316 eV),
may be traced to the granular nature of the disorder. Although, on
the average, the QDs are uniformly distributed, there exist
configurations with several dots coming close to each other, within
a distance of a dot size. Such rare configurations constitute more
efficient traps than the white-noise fluctuations and therefore
dominate in the deeper tail. The clusters of QDs are akin to
clusters of potential wells in the Lifshitz model of a disordered
electronic system~\cite{ref4a,ref4b}. Thus, anisotropic QD
clustering, can account for the shape and multiplicity of the
strongly localized modes in the deeper tail. We are unable to make
quantitative predictions for the density and spatial structure of
these modes, since no comprehensive theory
exists. Yet, we show that dielectric constant enhancement of order
$1\%$ over distance of order $1\mu m$, does account for the
experimental observations (Fig.~\ref{fig:two}e).

In conclusion, we demonstrate disorder-induced trapped photonic
modes in a microcavity with an embedded layer of QDs. In the lateral
in-plane direction the modes are localized by spatial fluctuations
of the dielectric constant due to randomness in the location and
composition of the QDs. Our theory emphasizes the universal features
of the trapping mechanism: the necessary combination of mirrors in
the vertical direction and the lateral disorder. The theory
quantitatively estimates the characteristic energy by which the PL
intensity decays below the PMC mode. Since the non universal,
system-specific details of the randomness are not accurately known,
the spatial shapes and energies of the
modes deeper in the energy tail, are qualitatively discussed only.

In addition we show both experimentally and theoretically that if
the size of the excitation spot is sufficiently small, as it is in
confocal and near field scanning microscopy, the microcavity
emission spectrum becomes very narrow, in marked difference with the
prediction of the standard cavity considerations, which apply to
uniform excitation. The disorder broadens this spectrum, as well.
Thus, we quantitatively measure and calculate the disorder induced
broadening below and above the cavity mode frequency.

\begin{acknowledgments}
This research is supported
by the German Israel and by the Israeli Science Foundations (GIF
and ISF) and by RBNI at the Technion.
\end{acknowledgments}

\end{document}